\documentclass[aps,preprint]{revtex4}
\usepackage{psfig}

\begin{document}
                                                                                         
\title{Design of HIV--1--PR inhibitors which do not create resistance: blocking the folding of single monomers}
\author{R. A. Broglia$^{1,2,3}$, G. Tiana$^{1,2}$, L. Sutto$^{1,2}$, D. Provasi$^{1,2}$ and F. Simona$^{1,2}$}
\address{$^{1}$Dipartimento di Fisica, Universit\'a di Milano, via Celoria 16, 20133 Milano, Italy}
\address{$^{2}$INFN, Sez. di Milano, Milano, Italy}
\address{$^{3}$Niels Bohr Institute, University of Copenhagen, Bledgamsvej 17, 2100 Copenhagen, Denmark}
\date{\today}

\begin{abstract}
One of the main problems of drug design is that of optimizing the drug--target interaction. In the case in which the target is a viral protein displaying a high mutation rate, a second problem arises, namely the eventual development of resistance. We wish to suggest a scheme for the design of non--conventional drugs which do not face any of these problems and apply it to the case of HIV--1 protease. It is based on the knowledge that the folding of single--domain proteins, like e.g. each of the monomers forming the HIV--1--PR homodimer, is controlled by local elementary structures (LES), stabilized by local contacts among hydrophobic, strongly interacting and highly conserved amino acids which play a central role in the folding process. Because LES have evolved over myriads of generations to recognize and strongly interact with each other so as to make the protein fold fast as well as to avoid aggregation with other proteins, highly specific (and thus little toxic) as well as effective folding--inhibitor drugs suggest themselves: short peptides (or eventually their mimetic molecules), displaying the same amino acid sequence of that of LES (p--LES). Aside from being specific and efficient, these inhibitors are expected not to induce resistance: in fact, mutations which successfully avoid their action imply the destabilization of one or more LES and thus should lead to protein denaturation. Making use of Monte Carlo simulations within the framework of a simple although not oversimplified model, which is able to reproduce the main thermodynamic as well as dynamic properties of monoglobular proteins, we first identify the LES of the HIV--1--PR and then show that the corresponding p--LES peptides act as effective inhibitors of the folding of the protease which do not create resistance.
\end{abstract}

\maketitle

\section{Introduction}

Because human immunodeficiency virus type--1 Protease (HIV--1--PR) is an essential enzyme in the viral life cycle, its inhibition can control acquired immune deficiency syndrome (AIDS).

The main properties inhibitory drugs must display are efficiency and specificity. Conventionally, this is achieved by either capping the active site of the enzyme (competitive inhibition) or, binding to some other part of the enzyme, by provoking structural changes which make the enzyme unfit to bind the substrate (allosteric inhibition). All the inhibitors of the HIV--1--PR available in the market (Indinavir, Sanquinavir, etc.) and which have been approved by FDA follow the former paradigm.

The large production of virions in the cell, coupled with the error prone replication mechanism of retroviruses, lead to escape mutants, drug resistance, and eventually persistance of the disease. Under the selective pressure of drugs, HIV--1--PR either mutates at the active site or at sites controlling its conformation, in such a way that the enzymatic activity is essentially retained, while the drug is not able to bind to its target anymore. The first signs of the failure of the drug usually takes place 6--8 months after the starting of the treatment \cite{tommasselli}.

We wish to suggest a novel type of HIV--1--PR inhibitor which interfere with the folding mechanism of the protein, destabilizing it and making it prone to proteolysis. These drugs are expected to be, aside from highly specific, perdurably efficient. In fact, as we shall see in the following, drug induced mutations will necessary affect sites important for the folding and for the stability of the protease, and consequently lead to its denaturation.

\begin{figure}
\centerline{\psfig{file=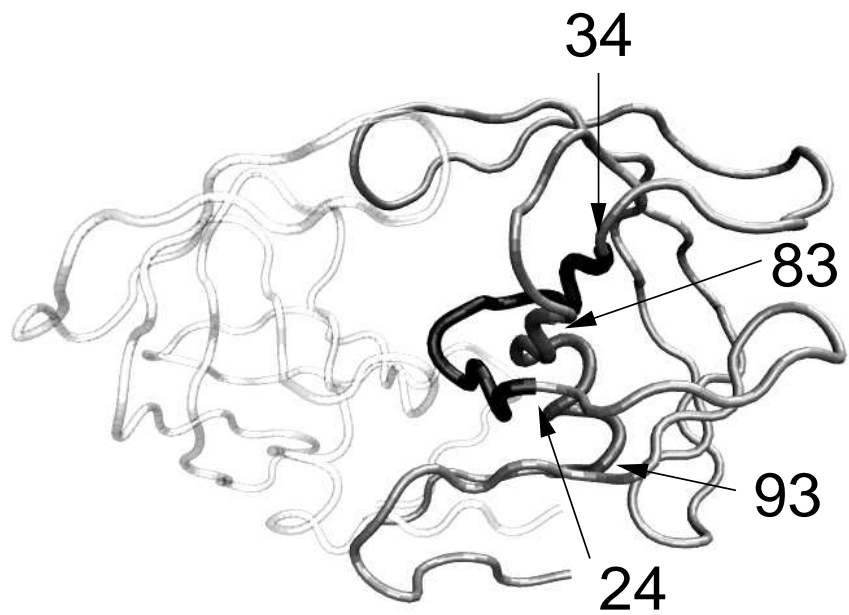,width=12cm}}
\caption{The native homodimer of HIV--1 PR. In the right monomer we have highlighted the local elementary structures, corresponding to fragments 24--34 and 83--93 (see text).}
\label{fig_native}
\end{figure}

HIV--1--PR is a homodimer (cf. Fig. \ref{fig_native}), that is a protein whose native conformation is built out of two (identical) disjoint chains. Sedimentation equilibrium experiments have shown that, in a neutral solution (pH=7, $T=4^0$C), the protease folds according to a three--state mechanism ($2U\rightarrow 2N\rightarrow N_2$), populating consistently the monomeric native conformation $N$ \cite{xie}. This result (cf. also the Appendix \ref{appA}) is supported by NMR studies of mutants where the interaction across the interface is weakened but the monomer retain its native conformation \cite{ishima}, by all--atom simulations of the HIV--PR monomer in explicit solvent \cite{caflisch} and by G\=o--model simulations of the dimer \cite{levy}. The dimer dissociation constant ($2N\leftrightarrow N_2$) is found to be $k_d=5.8\mu M$ at $T=4^0$C \cite{xie}: for instance, in a 30 $\mu$M solution 44\% of proteins are in monomeric form. This allows one to conclude that, at neutral pH, each monomer of the protein folds following the same hierarchical folding mechanism of single domain, monomeric proteins \cite{dimer}: after the monomer has reached the native state, it diffuses to find another folded monomer to associate with.

A number of experimental and theoretical evidences suggests that globular, single--domain proteins avoid a time--consuming search in conformational space, folding through a hierarchical mechanism. Ptitsyn and Rashin observed a hierarchical pathway in the folding of Mb \cite{ptytsin}. Lesk and Rose identified the units building the folding hierarchy of Mb and RNase on the basis of geometric arguments \cite{lesk}, deriving the complete tree of events which lead these proteins to the native state.  All these studies describe a framework where small units composed of few consecutive amino acids build larger units which, in turn, build even larger ones, which eventually involve the whole protein \cite{baldwin}. The kinetic advantage of this mechanism is that, at each level of the hierarchy, only a limited search is needed for the smaller units to coalesce into the larger units belonging to the following level \cite{panchenko}.

Lattice model calculations \cite{jchemphys2,jchemphys3} have shown that the folding of a small monomeric protein proceeds, starting from an unfolded conformation, follow a hierarchical succession of events: 1) formation of local elementary structures (LES, containing 20\%--30\% of the proteins amino acids) stabilized by few highly conserved, strongly--interacting ("hot") hydrophobic amino acids ($\leq 10\%$ of the proteins amino acids) lying close along the polypeptide chain, 2) docking of the LES into the (postcritical) folding nucleus \cite{sh_nucleo}, that is formation of the minimum set of native contacts which brings the system over the major free energy barrier of the whole folding process, 3) relaxation of the remaining amino acids on the native structure shortly after the formation of the folding nucleus. The "hot" sites, which stabilize the LES are found to be very sensitive to (non--conservative) point mutations. Since most of the protein stabilization energy is concentrated in these sites, mutating one or two of them has a high probability of denaturing the native state. On the other hand, mutating any other site ("cold" sites, even those "cold" sites belonging to the LES) has in general little effect on the stability of the protein \cite{aggreg,tiana98}.

Making use of the same model it has been shown that it is possible to destabilize the native conformation of a protein making use of peptides whose sequence is identical to that of the protein LES \cite{rudi}. Such peptides interact with the protein (in particular with their complementary fragments in the folding nucleus) with the same energy which stabilizes the nucleus, thus competing with its formation. 

There are two important advantages of these folding--inhibitors with respect to conventional ones. First, their molecular structure is suggested directly by the target protein. One has not to design or to optimize anything, just find which are the LES of the protein that has to be inhibited. The design has been performed by evolution through a myriad of generations of the virus (or of the organism which expresses the protein). Moreover, it is unlikely that the protein can develop resistance through mutations. In fact, the present inhibitor binds to a LES, and a protein cannot mutate a LES \cite{aggreg}, in any case not those "hot" amino acids which are essential to stabilize it as well as to bind to the other LES to form the folding nucleus \cite{tiana98}, under risk of denaturation. Note that, within this context, neutral mutations (e.g., hydrophobic--hydrophobic) of these hot amino acids are possible, as they do not essentially change the stability of the corresponding LES, nor the strenght and specificity with which LES dock to form the folding nucleus.

\section{Model calculations of the folding properties of HIV--1 PR} \label{sect_model}

The investigation of the folding of the HIV--1 PR has two goals: i) to validate the model which will be employed to study the effect of the inhibitor peptides and ii) to help locate the local elementary structures of the protein.  For this purpose, use is made of a modified G\=o model. In standard G\=o model calculations \cite{go}, as that carried out by Levy and coworkers \cite{levy} in their study of the HIV--1--PR, the interaction between each pair of amino acids is described by a square well whose bottom lies at the same energy for all native pairs and at zero or positive energy for non--native pairs. Such a treatment insures the native conformation to be the global energy minimum of the system, providing at the same time a realistic description of the entropic features of the chain. It however fails in providing any chemical characterization of the amino acids, treating all of them on equal footing. To account for the diversity existing between amino acids, we assign to each native pair an interaction energy obtained averaging Gromacs \cite{gromacs} force field around the native conformation of the monomer. This procedure has proven useful to account for the folding properties of a number of small, single--domain proteins \cite{ludo}. 

The model pictures each amino acid as a spherical bead, making inextensible links with the following one. The amino acids interact through a contact potential of the kind
\begin{eqnarray}
U(\{r_i\})&=&\sum_{i+2<j} \left[ B_{ij}\theta(R-|r_i-r_j|)\theta(R-|r_i^N-r_j^N|)+\epsilon_{HC}\theta(0.99\cdot|r_i^N-r_j^N|-|r_i-r_j|)\times \right. \nonumber\\
  &\times&\left.\theta(R-|r_i^N-r_j^N|)+\epsilon_{HC}\theta(R-|r_i-r_j|)[1-\theta(R-|r_i^N-r_j^N|)]\right]
\label{potential}
\end{eqnarray}
where $r_i$ is the coordinate of the C$_\alpha$ atom of the $i$th amino acid, $r_i^N$ is the coordinate in the crystallographic native conformation (pdb code 1BVG), $\theta(x)$ is a Heaviside step function, $B_{ij}$ is the interaction energy between $i$th and $j$th amino acid, $R$ the interaction range (which in the following calculations is set equal to $7.5\AA$) and $\epsilon_{HC}$ is the hard core repulsion, set to $100\;kT$. Accordingly, the first term of the potential function describes the attraction between native pairs, the second term the hard core between native pairs, its range being equal to the 99\% of the native distance, and the last term describes the repulsion between non--native pairs. Moreover, we assume that residue $i$ interacts with residue $i+2$ only through a hard core repulsion of range $R/2$. 

To determine the quantity $B_{ij}$ all atom molecular dynamic simulations were carried out making use of the Gromacs package, treating explicitely the solvent. The simulations were done for $1$ ns at room temperature around the native conformation of the dimer. During this time interval the overall RMSD of the system did not exceeded $2.5\AA$. The values $B_{ij}$ are the result of the average over the full simulation of the interaction energies between the different pairs of amino acids.  

\begin{figure}
\centerline{\psfig{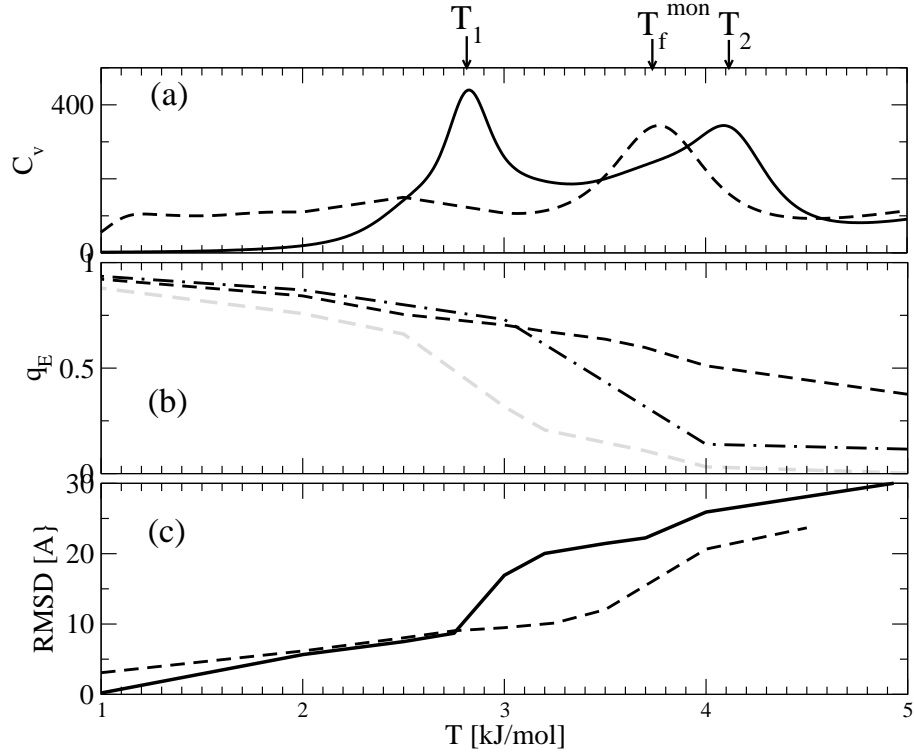}}
\caption{(a) The specific heat of the dimer (solid curve, the values obtained from the simulations have been divided by 2 in order to obtained the specific heat {\it per monome}) and of the monomer (dashed curve); (b) the order parameter $q_E$ associated with the contacts within monomers (dashed black curve), across the dimer (dashed gray curve) and within the nucleus (contacts between fragment 22--32 and 83--93) of the monomer (dot--dashed black curve); (c) the RMSD associated with the whole dimer (solid curve) and with the monomer alone (dashed curve).}
\label{cv}
\end{figure}

The simulations carried out making use of the modified G\=o model were performed in a $100\AA$ box with periodic boundary conditions (equivalent to a 10 mM concentration), and  temperatures ranging from 1 to 5 kJ/mol (in the following, temperatures will be expressed in kJ/mol, setting Boltzmann's constant equal to 1: for instance, 300K correspond to $\approx$2.5 kJ/mol). From these simulations,characterization of the dimer thermodynamical quantities have been obtained by means of a modified multi--histogram techniques \cite{borg}. The resulting dimer specific heat (per monomer) is displayed in Fig. \ref{cv}(a) (solid curve). Consistent with the findings of ref. \cite{levy}, it displays two peaks, one at $T_1=2.8$ kJ/mol and one at $T_2=4.1$ kJ/mol. 

To better quantify the properties of the protein we introduce the parameter $q_E$ defined as the fraction of the native energy (e.g., $q_E=1$ means that the dimer is in the native conformation). We will use the parameter $q_E$ also with respect to the monomer or to the interface, to indicate the fraction of energy within the monomer or at the interface of a given conformation.  In Fig. \ref{cv}(b) is displayed the value of the parameter $q_E$ associated with the interaction within each of the monomers forming the HIV--1--PR (black dashed curve),as well as that associated with the dimerization, that is the interaction between the monomers (dashed gray curve). The decrease in the interaction energy at the interface between the two monomers taking place at $T_1$ indicates that the associated peak in the specific heat of the two chains (continuous curve in Fig. \ref{cv}(a)) marks the transition between the dimeric and the monomeric forms of the protein. Just above $T_1$, each monomer is still in the native basin of attraction, displaying a $q_E\approx 0.75$. The temperature $T_2$ corresponds to the (weak) transition to unfolded monomers ($q_E<0.5$, cf. dashed black curve in Fig. \ref{cv}(b)).

The same kind of weak transition is found for simulations of an HIV--1--PR monomer alone (dashed curve in Fig. \ref{cv}(a)), although at a slightly lower temperature ($T_f^{mon}=3.8$ kJ/mol). The weakness of the (monomer) folding transition (taking place at $T_2$) is associated with a faint degree of cooperativity, as testified by the low value assumed by the two--state parameter \cite{kaya} $\kappa_2=0.18$. This parameters ranges from $1$ for fully cooperative transitions, to $0$ for non--cooperative transitions. Although it is well known that simplified models understimate the cooperativity of the folding transition \cite{thirumalai}, the HIV--1--PR monomer displays a value of $\kappa_2$ which is much lower than that of other proteins simulated with the same model (e.g., src--SH3 displays $\kappa_2=0.38$, even if it is shorter than the HIV--1--PR).

The physical reason why the folding of the HIV--1--PR monomer is much less cooperative than other monoglobular proteins can be found on the properties of the folding nucleus of each of the monomers forming this protein. As discussed below, the folding nucleus of the protease is built out of the LES containing the monomers 24--34, 83--93 and 75--78. Due to the distance between the two LES along the chain, the assembly of the folding nucleus leaves a conformational freedom to the rest of the chain (specifically, to the fragment 35--75 at least and likely also to the fragment 35--83) uncommon among other proteins. This is testified by the large equilibrium RMSD found under folding conditions (up to $10\AA$ for $T< 3.8$ kJ/mol) in simulations of the  monomer alone (dashed curve in Fig. \ref{cv}(c)), where the folding nucleus is essentially formed (dashed--dotted black curve in Fig. \ref{cv}(b)). Within this context, we note that the basin of attraction of the monomeric native state extends to conformations with RMSD of 10$\AA$, correspondindg to a typical $q_E$ of 0.7 (cf. Fig. \ref{fig_free3d}(a)), while the unfolded states has a RMSD of the order of 12$\AA$ and beyond. The large fluctuations of the fragment 35--75 (or 35--83) when the nucleus is formed produce a shoulder in the specifc heat at low temperatures (cf. dashed curve in Fig. \ref{cv}(a)) and blurr the folding transition at $T_f^{mon}$.

\begin{figure}
\centerline{\psfig{file=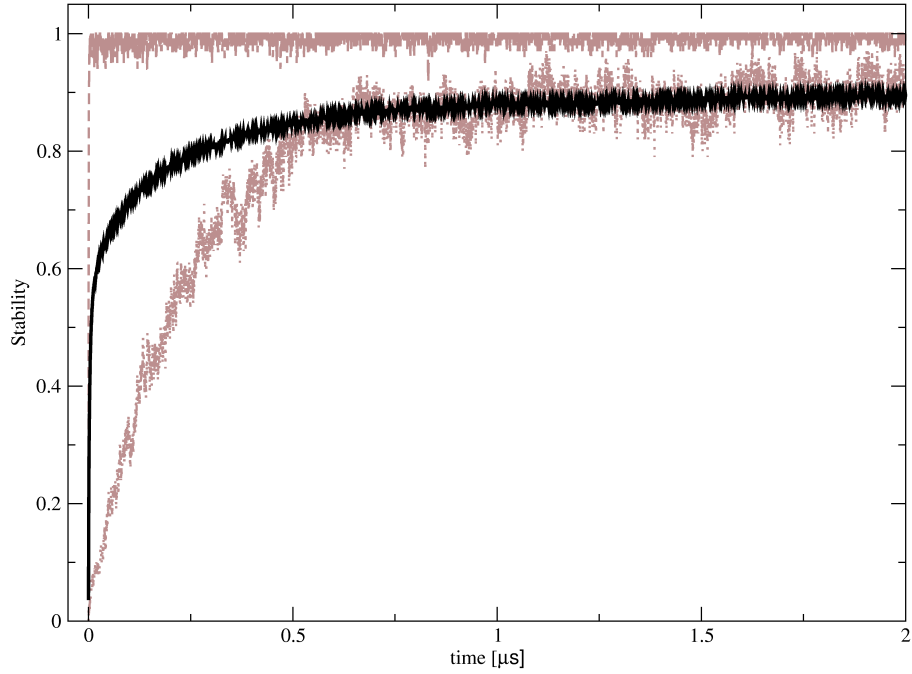,width=12cm}}
\caption{Dynamic evolution of the monomer. The formation probability $[q_E](t)$ of the native conformation (solid curve) at $T=2.5$ kJ/mol, the probability $p_{87-90}(t)$ (dashed gray curve) and the probability $p_{31-89}(t)$ (dotted gray curve).}
\label{fig_dyn}
\end{figure}

With the same model it is possible to run dynamical simulations, starting from random conformations and follow the folding of the protein into the native state. Since we are interested in inhibiting the folding of the HIV--1--PR monomer, we will concentrate on the dynamics of the monomer. The overall dynamics can be followed through the plot of $[q_E](t)$, that is the fractional native energy as a function of time, averaged over 100 independent runs. The result at $T=2.5$ kJ/mol is displayed as a solid curve in Fig. \ref{fig_dyn}, indicating an exponential process of characteristic time $\tau=2.9\cdot 10^{-7}$s (cf. Appendix B), consistent with the two--state picture. The model provide also information concerning the formation of each contact, through the probability $p_{ij}(t)$ that the contact between residues $i$ and $j$ is formed at time $t$. A number of contacts are stabilized early (sub--nanosecond time scale) following an exponential dynamics. This is the case of, for example, contact 87--90. Contacts between residues which are far along the chain are formed later, following a non--exponential dynamics, which indicate that their formation is dependent on some other event\cite{ludo}. The earliest among them involve the fragments 22--34 and 77--93 after an average time $\tau\approx 10^{-7}$ s. As an example, in Fig. \ref{fig_dyn} is displayed the formation probability of the contact 31--89. 

\begin{figure}
\centerline{\psfig{file=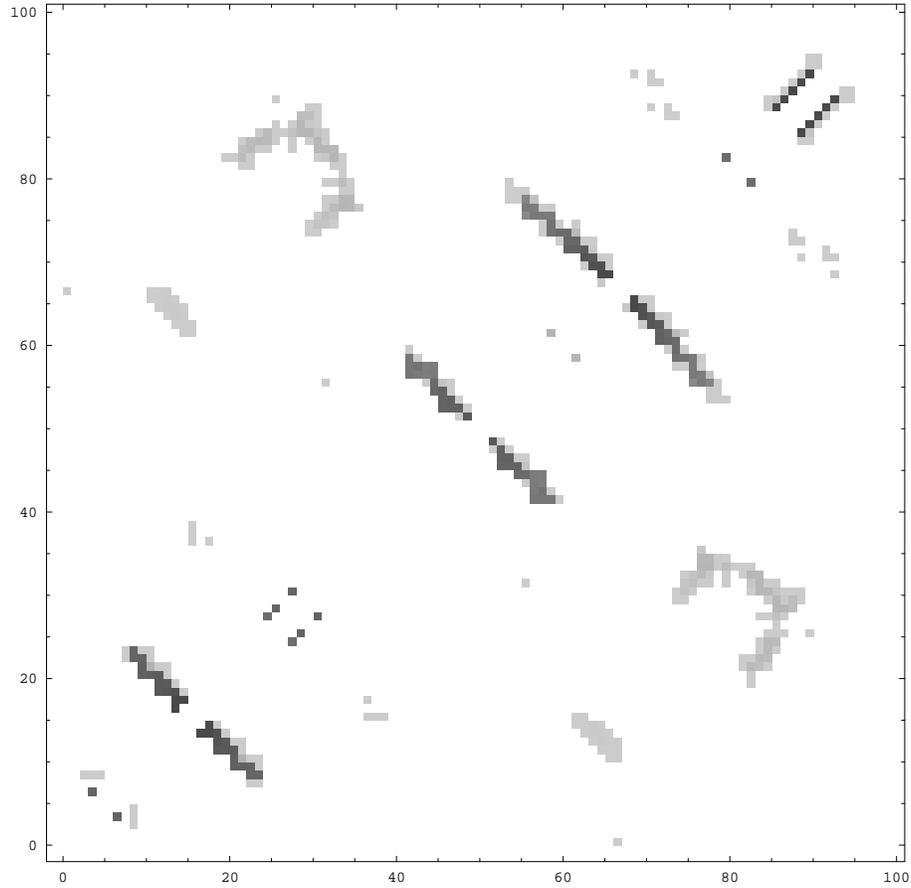,width=12cm}}
\caption{The contact map of HIV--1--PR, where different gray levels correspond to different values of the average stabilization time. Darker and lighter symbols correspond to times of the order of $10^{-10}$s and $10^{-7}$s, respectively.}
\label{fig_dyn2}
\end{figure}

In Fig. \ref{fig_dyn2} is summarized the hierarchy of formation of native contacts of HIV--1--PR, the different gray levels corresponding to different time scales, while in Table \ref{table_dyn} are listed the parameters associated with selected contacts. The picture that emerges is that local contacts within fragments 83--93 and in the beta--hairpin 42--58 form first. Note also the very fast formation of contact 25--28 belonging to the fragment 22--34. Then the beta--turns 14--19 and 64--72, again built out of local residues. The next event is the assembly of the nucleus involving fragments 22--34 and 83--93, which is further stabilized by the contribution of the strongly--interacting segment 75--78. Finally, the rest of the residues come to place.

Summing up, the model suggests that LES are built of residues which lie in the regions 83--93, 24--34 and likely also 75--78. This result essentially agrees with the indications provided by the studies of Levy and coworkers \cite{levy}. These authors have found that the group of amino acids 27--35 and 79--87 are essential in the folding of the protein.

\section{Localization of the LES: indirect methods} \label{sect_les}

The direct inspection of the dynamics of the HIV--1 PR obtained by means of model simulations of the folding have suggested which are the LES controlling the folding of the monomer. It can be interesting to study other techniques to localize the LES, in order both to substantiate the findings of the model, and to develop a more economical method to be used in the case of other proteins one wishes to inhibit.

\subsection{Evolutionary data}

Since LES are responsible for guiding a protein to its native conformation, it is likely that evolution pays particular care in conserving their sequence. In fact, model simulations have shown that residues building the LES can only undergo conservative point mutations (e.g., hydrophobic--hydrophobic), at the risk of denaturing the protein \cite{tiana98}. Consequently, LES are highly conserved in families of structurally similar proteins \cite{isles}. Comparative studies of a number of protein families have shown that this is indeed the case \cite{mirny99}. 

\begin{figure}
\centerline{\psfig{file=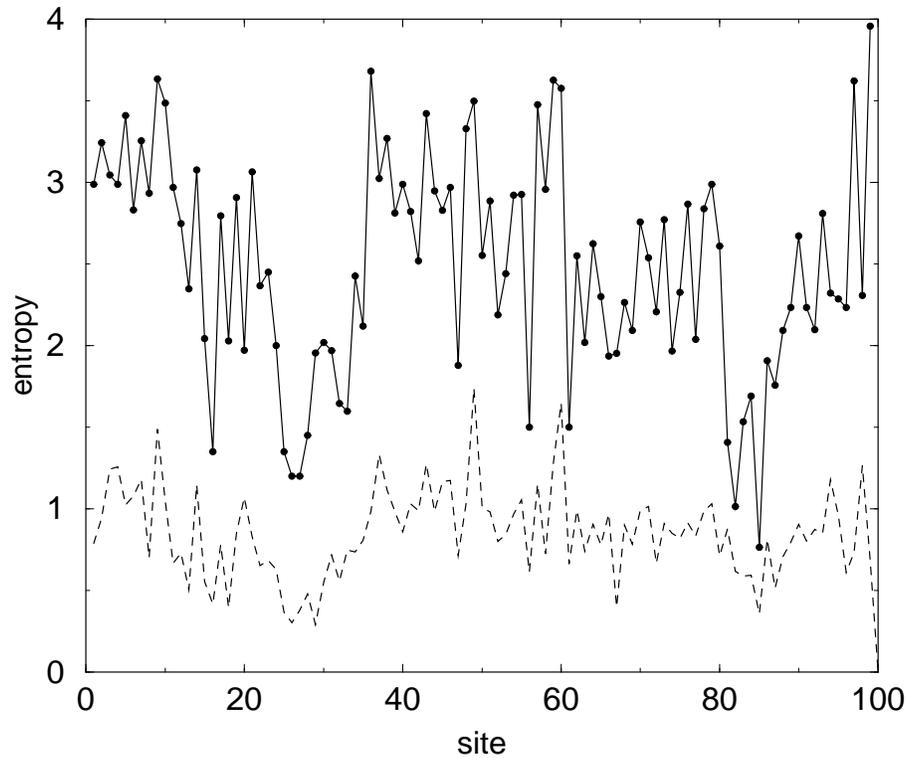,width=12cm,angle=-90}}
\caption{The entropy per site (see text) of proteins structurally similar to the HIV--1--PR monomer (pdb code: 1BVG). The solid line indicates the entropy function calculated over 28 proteins displaying sequence similarity lower than 25\% with the HIV--1--PR, while the dashed line is associated with 462 proteins irrespectively of sequence similarity.}
\label{fig_entropy}
\end{figure}

A measure of the degree of conservation of residues in a family of proteins is provided by the entropy per site $S(i)\equiv -\sum_\sigma p_i(\sigma)\log p_i(\sigma)$, where $p_i(\sigma)$ is the frequency of appearence of residue of type $\sigma$ at site $i$ in the proteins belonging to the family. In order to be statistically meaningful, we have plotted in Fig. \ref{fig_entropy} (with a solid line), the entropy calculated over a family of 28 uncorrelated proteins (i.e., sequence similarity lower than 25\%), structurally similar to HIV--1--PR \cite{fssp}. As seen from the figure, the most conserved regions are those involving residues 22--33 and 81--90. Even disregarding the statistical caution and calculating the entropy over 462 proteins \cite{hssp} displaying any sequence similarity to HIV--1--PR (dashed line in Fig. \ref{fig_entropy}) the plot indicates the same regions as the most conserved. Note that the conservation of residues 25--27 is anyway not unexpected, in that they build the active site of the protease (D25, T26 and G27).

Another important source of information concerning the HIV--1--PR is provided by the study of its sequences in specimens coming from infected individuals.  Being a retrovirus, HIV can replicate very fast but very imprecisely, thus displaying a rather fast evolution rate. This evolution is reflected by the appearence of a manyfold of mutated HIV--1--PR which retain their folding features. In Table \ref{table_mut} are listed the mutations observed in  28417 isolates coming from patients infected with HIV--1--PR \cite{shafer}. Since some mutations can be conservative, that is substitute an amino acid with another displaying similar chemical properties, we also display on the table the lowest PAM250 \cite{pam250} score associated with the mutations in each site. A positive PAM250 score indicates a conservative mutation. Although the fragments with no mutations or with only conservative mutations are too many to let one identify the LES from this information alone (probably because of the limited statistics in the database), the fact that only conservative mutations fall in the fragments 24--34, 83--93 and 75--78 is consistent with the description of the folding nucleus made above.
  
\subsection{Static energy features} \label{sect_energy}

\begin{figure}
\centerline{\psfig{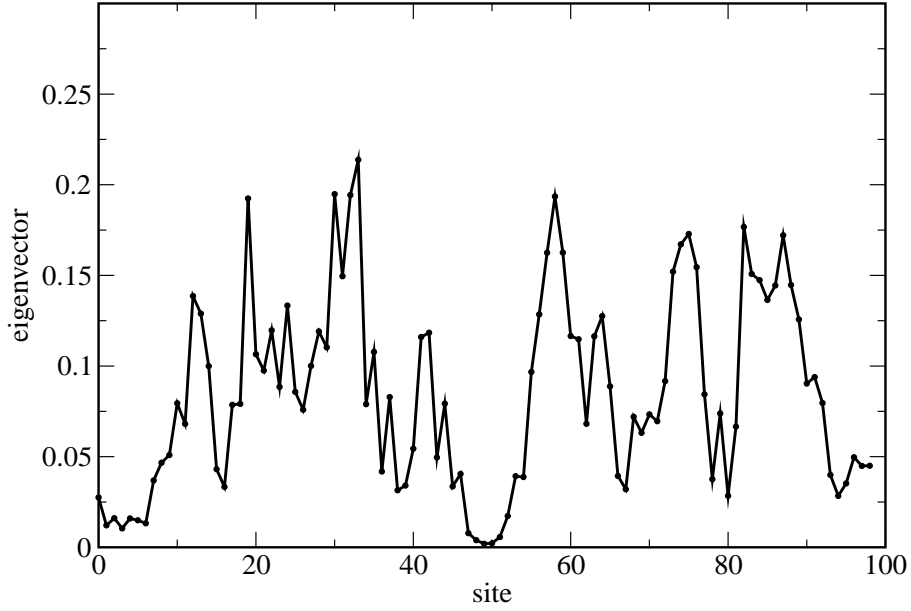}}
\caption{The components of the eigenvector associated with the lowest eigenvalue of the interaction matrix $B_{ij}$.}
\label{fig_eigen}
\end{figure}

In order to become stable at an early stage of the folding process, LES must carry a significant fraction of the total energy of the protein in its native conformation. We have analyzed the native interaction $B_{ij}$ between the amino acids following the scheme described in ref. \cite{giorgio}, through an eigenvalue decomposition of the matrix. The lowest energy state ($\lambda_1$=-121.2 kJ/mol) displays a large energy gap (i.e. -12.9 kJ/mol$\approx 5kT$) with respect to the next eigenvalue, indicating a core of strongly interacting amino acids. We display in Fig. \ref{fig_eigen} the eigenvector associated with the lowest energy eigenvalue, which highlights to which extent the different amino acids participate to this core. The largest amplitudes involve residues 25--34, 57--65, 75--77 and 83--90. These regions of residues overlap well with the conserved regions mentioned above in connection with Fig. \ref{fig_entropy}.

Another way of representing the interaction matrix $\parallel B_{ij}\parallel$ is to consider the interaction between fragments $S_1=(13-21)$, $S_2=(24-34)$, $S_3=(38-48)$, $S_4=(50-55)$, $S_5=(56-66)$, $S_6=(67-72)$, $S_7=(75-78)$, $S_8=(83-93)$. The corresponding $8\times 8$ energy map is essentially codiagonal, the associated energy of the interacting chain segments being $-1936$ kJ/mol as compared to the MD simulation native energy $-2722$ kJ/mol. In other words, the $S_1-S_8$ representation of the folded monomer accounts for $\approx 70\%$ of the calculated native conformation energy. Because this representation contains 70 residues, it would be tempting to conclude that the native energy is uniformly distributed over all the amino acids. Within this context, it is useful to calculate the difference between the native S$_1$--S$_8$ energy map and that associated with the unfolded (U) state ($q<0.3$). For this purpose, G\=o model simulations (cf. Sect. \ref{sect_model} as well as discussion below) have been carried out to obtain a statistically representative ensemble of U--states, from which we have extracted an average set $\{q_i\}_U$ of similarity parameters. Weighteing each contribution to the elements of the $8\times 8$ matrix by the corresponding difference $(q_N-q_U)_i$, one obtains the energy map shown in Table \ref{table1}.
It is seen that the $S_1-S_8$ representation divides into three blocks: a) one composed of segments S$_2$, S$_7$ and S$_8$, b) one composed of segments S$_1$ and S$_6$, and c) one containing segments S$_3$, S$_4$ and S$_5$.
While the average energy per residue in those block is $-5.3$ kJ/mol, that associated with S$_2$+S$_7$+S$_8$ and S$_1$+S$_6$+S$_3$+S$_4$+S$_5$ is -8.3 kJ/mol and -4.9 kJ/mol, respectively.

The above results indicate that the $S_2$, $S_7$ and $S_8$ segments qualify as LES (folding units) of each of the two monomers of the HIV-1-PR dimer, LES which form in their native conformation the (post critical) folding nucleus (FN). It is interesting to note that drug induced mutations in the amino acids belonging to these LES (L24I, D30N, L33F, V77I, I84V, I85V, N88D and L90M \cite{tommasselli}), lead to a folding nucleus energy equal to -1500 kJ/mol, as compared to -796 kJ/mol for the wild type sequence FN, that is to an increase of almost a factor of 2 in the stability of the system, considering the effect of all mutations simultaneously. In fact, single selected mutations add to the FN stability 5-10 kJ/mol. Within this context and that of Fig. \ref{fig_eigen} one can identify sites 33, 75, 76, 84 and 89 as "hot" sites.

\subsection{Further evidence}

Wallqvist and coworkers \cite{wallqvist} investigated the HIV--1--Protease molecule for the occurence of cooperative folding units that exhibit a relatively stronger protection against unfolding than other parts of the molecule. Unfolding penalities are calculated forming all possible combinations of interactions between segments of the native conformation and making use of a knowledge--based potential. This procedure identifies a folding core in HIV--1--PR comprising residues 22--32, 74--78 and 84--91, residues that form a spatially close unit of a helix (84--91) with sheet (74--78) above another $\beta$--strand ((22--25), containing the active site residues D25, T26 and G27) perpendicular to these elements.

Making use of a Gaussian network model, Bahar and coworkers \cite{bahar} studied the normal modes about the native conformation of HIV--1--Pr. "Hot" residues, playing a key role in the stability of the protein, are defined as those displaying the fastest modes. In this way are identified regions 22--32, 74--78 and 84--91 as folding core of the protein. These regions match with those displaying low experimental Debye-Weller factors, that is low fluctuations in the crystallographic structure.

Calculation of $\varphi$--values by means of G\=o--model simulations performed by Levy and coworkers \cite{levy} have located a major transition state where only regions 27--35 and 79--87 are structured. The protein then reaches the native state, overcoming another minor transition state, and subsequently dimerizes into the biologically active structure.

Cecconi and coworkers \cite{cecconi} have calculated the stability temperatures associated with each contact of the protease, again making use of a G\=o model. They find that key sites to the stability of partially folded states, that is those displaying the lowest stability temperatures, are 22, 29, 32, 76, 84 and 86.

\section{Inhibition of HIV--1 PR} \label{sect_inh}

The central issue of the present work is to show that it is possible to destabilize the native state of the HIV--1 PR monomer, shifting the equilibrium to the unfolded state, by means of short peptides displaying the same sequence as one of the LES (which we shall call peptides p--LES).

As emerged from the calculations discussed in Sect. \ref{sect_model} and the evidence presented in Sect. \ref{sect_les}, the segments 24--34 (S$_2$), 83--93 (S$_8$) and likely 75--78 (S$_7$) qualify as LES of the HIV--1--PR monomer, and thus as leads of inhibitors of the enzyme, with the following provisos. Segment S$_7$ is a so called open LES (cf. ref. \cite{1d3d}), too short to be specific. Concerning the S$_2$ LES, it contains the active site (residues 25, 26 and 27). In the model studies of the design of good folders, neither we nor any other group has ever considered the role the conserved amino acid belonging to the active site play in the resulting sequence, nor in  the folding properties of the protein. Nonetheless one knows that this role is likely to be important. In particular, while considerations of hydrophobicity and/or  capability to establish the largest number of native contacts suggests that the most strongly interacting amino acids providing the stabilization of the LES and thus of the folding nucleus should be, in the native conformation, well protected and buried inside the protein, those associated with the active site should be reachable by the substrate and thus lay on the surface of the enzyme. The corresponding frustration is well exemplified by the anticorrelation observed between the entropy values associated with sites 26 and 27 (low) and the corresponding eigenvector components (also low) shown in Figs. \ref{fig_eigen} and \ref{fig_entropy}. This anticorrelation becomes even stronger if one compares it with the perfect correlation existing between the values of the entropy (low) and of eigenvector (high) associated with sites 33 and 85 essential in the folding process (hot sites) but not connected with the active site. Summing up, we do not know what the consequences are of this frustration in the design of nonconventional inhibitors. Consequently, in what follows we shall exclusively concentrate on the LES S$_8$, and on the inhibitory properties the peptide p--S$_8$ has.

\begin{figure}
\centerline{\psfig{file=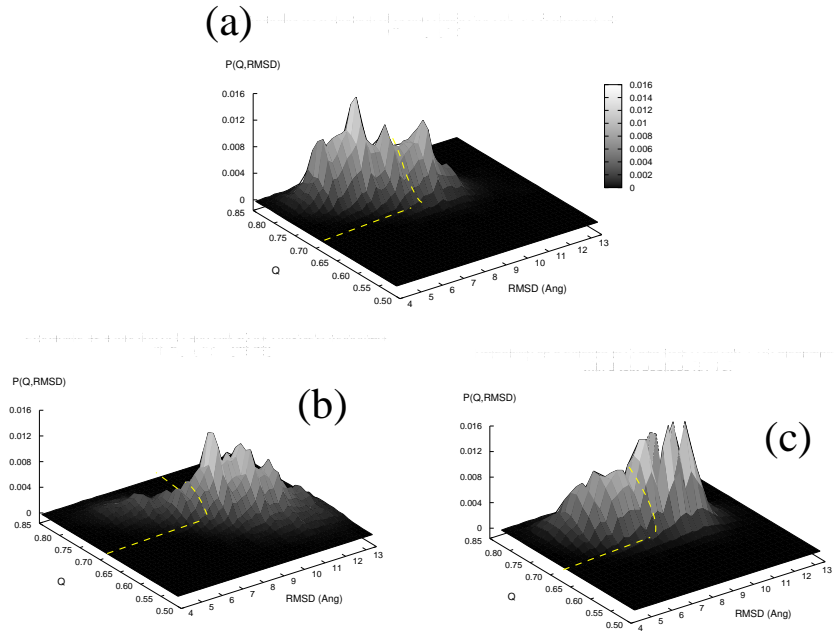,width=12cm}}
\caption{The equilibrium probability of the HIV--1 PR monomer as a function of the energetic parameter $q_E$ and of the RMSD for the monomer alone (a), the system composed of the monomer and 3 peptides p--S$_8$ (b) and of the monomer and 3 peptides corresponding to the sequence 5--15 control peptide (c). The dashed curve indicates the native state.}
\label{fig_free3d}
\end{figure}

For this purpose, we have performed equilibrium simulations of the system composed of the HIV--1 PR monomer and a number of p--LES (p--S$_8$) corresponding to the fragment 83--93 of the protein (S$_8$). The joint probability distribution $p(q_E,RMSD)$ of the native relative energy fraction $q_E$ and of the RMSD (normalized, i.e., divided by the number of residues) for the case of 3 p--LES at $T=2.5$ kJ/mol is displayed in Fig. \ref{fig_free3d}(b), to be compared with that of the monomer alone (Fig. \ref{fig_free3d}(a)). Consistently with the folding transition observed in Fig. \ref{cv} and with the discussion of Sect. \ref{sect_model}, we define as native state the region of Fig. \ref{fig_free3d}(a) characterized by $q_E>0.7$ and RMSD$<10\AA$ (this region is delimited with a dashed curve in the figure). The effect of the p--S$_8$ is to decrease drastically the population of the native state and increase at the same time that of the unfolded state. 

\begin{figure}
\centerline{\psfig{file=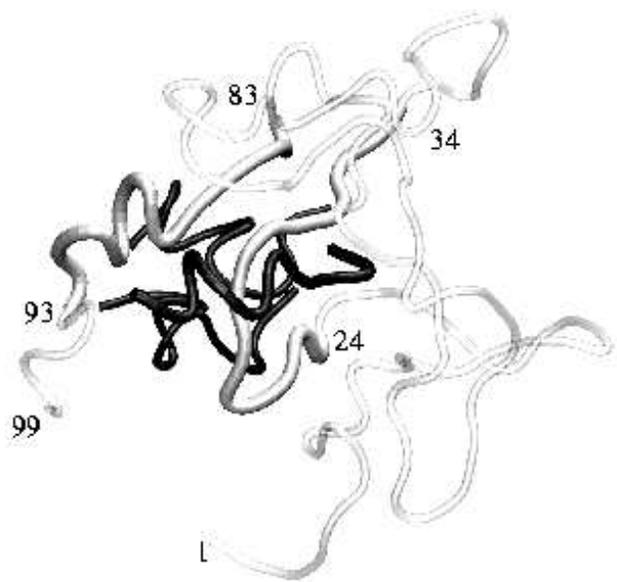,width=12cm}}
\caption{A snapshot of an unfolded conformation taken from the simulation of Fig. \protect\ref{fig_free3d}, where the peptides p--S$_8$ prevent the folding nucleus of the monomer to get formed.}
\label{snap}
\end{figure}

The increase of the peak associated with the unfolded state is caused by the appearence of conformations where the p--S$_8$ peptides are bound to the fragment 24--34 (S$_2$) of the protein, preventing the actual LES 24--34 to find its native conformation. An example of such conformation is shown in Fig. \ref{snap}, corresponding to the values $q_E=0.6$ and RMSD=$11\AA$. This conformation is particularly stable because the interaction between the p--LES and the monomer is of the order of $-165$ kJ/mol, the same amount of energy which stabilizes the nucleus of the protein. Note also that more than one p--S$_8$ is able to bind at the same time to the monomer, increasing the degree of denaturation.

\begin{figure}
\centerline{\psfig{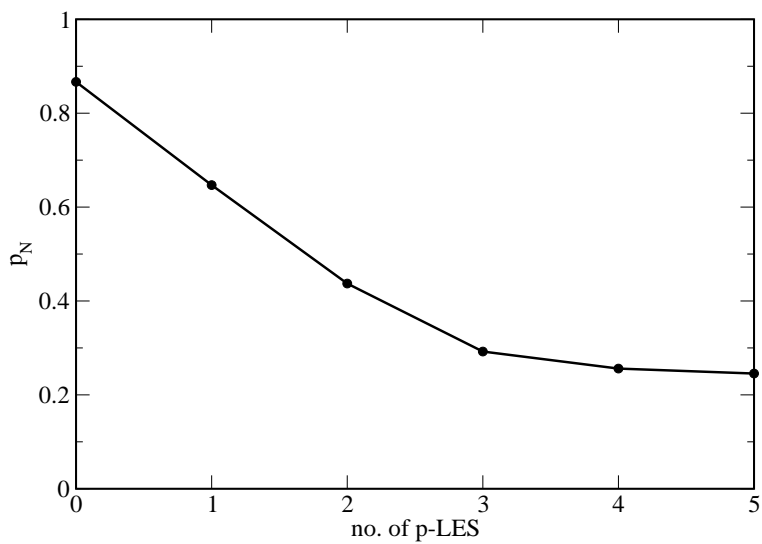}}
\caption{The equilibrium population ($p_N(\equiv p(q_E>0.7,\; RMDS<10\AA))$, see text Sect. \protect\ref{sect_inh}) of the native state of the monomer as a function of the number of peptides p--S$_8$ ($T=2.5$ kJ/mol).}
\label{p_t}
\end{figure}

In Fig. \ref{p_t} is displayed the equilibrium population $p_N$ of the native state as a function of the number $n_p$ of p--S$_8$ peptides. The inhibitory effect of the peptides is already present at $n_p=1$ (i.e., a concentration of p--S$_8$ equal to the concentration of protein, in terms of number of chains), in which case the stability of the native state is reduced by $\approx 30\%$ with respect to the situation with no peptides, while for $n_P=4$ the value of $p_N$ becomes $\approx 0.25$. We have repeated the same calculation with control peptides whose sequences are equal to that of the fragments 5--15 and 61--70 of the monomer. In all cases a slight decrease of the stability has been observed. However, the control peptides are not able to disrupt to any extent the folding nucleus and thus to prevent the monomer to reach the native state.

We have also calculated how inhibition depends on temperature, carrying out simulations of folding of the monomer protein in presence of three p--S$_8$ at two temperatures different from that used in connection with the results displayed in Fig. \ref{p_t}. At $T=3.5$ kJ/mol the value of $p_N$ drops to $0.01$, while at the (very) low temperature $T=2$ kJ/mol the simulation leads to $p_N=0.98$, indicating that p--S$_8$ becomes ineffective. Within this context, the following considerations are in place. At the "biological" temperature $T=2.5$ kJ/mol  our simulations are well equilibrated, in the sense that the system has gone in and out the native conformation a large number of times and the system has lost memory of the initial condition (e.g., simulations starting from the native conformation give the same results as simulations starting from a random, unfolded conformation). At $T=2$ kJ/mol  this is not true anymore. In fact, since in this case the dynamics is much slower, it is not feasable to perform simulations for long enough time so that the system becomes equilibrated. 

In any case, the low--temperature simulations performed give a strong signal about a diminished inhibitory effect of the p--S$_8$. This is not unexpected, due to the fact that the preference the protein has to bind a p--LES instead of its own LES is mainly entropical. More precisely, the protein can either make its S$_2$-- and S$_8$--LES interact to build the folding nucleus (in which case the protein folds) or make its S$_2$--LES interact with the p--S$_8$ peptide (in which case the protein does not fold). 
The difference between the two situations can thus be understood as follows. In the case where two LES bind to each other in the native conformation the whole system gains (potential energy) from the folding of the rest of the protein, paying at the same time the entropic cost associated with this phenomenon. In the case in which a LES binds a p--LES, the system essentially gains the same (potential) energy as in the previous case, due to the fact that most of the stabilization energy is concentrated in the interaction among the LES. On the other hand, the entropy cost is only that associated with the freezing out the degrees of freedom of the peptide. This entropic cost decreases with the number $n_p$ of p--LES in the system as $TS^{rot-transl}-T\log n_p$, the roto--translational entropy of the p--LES depending weakly on $n_p$ at low concentrations, as in the case under discussion. Consequently, regardless the stabilization energy of the protein monomer, there always exists a number $n_p$ of p--LES such that the free energy of the unfolded state is lower than that of the native state. At the "biological" temperature $T=2.5$ kJ/mol we observe (cf. Fig. \ref{p_t}) that $n_p=1$ is enough to destabilze the native state to a measurable extent. When the temperature is lowered, entropy plays a less important role (i.e., $F=E-TS$) and the equilibrium state becomes the lowest potential energy state, that is the native state.

The fact that the p--S$_8$ destabilizes the monomeric protease is a sufficient condition to prevent the replication of the virus. This in keeping with the fact that the monomer is at equilibrium with the dimer. Consequently, the destabilization of the monomer shifts the equilibrium of the system towards the unfolded state. Moreover, the fact that the monomeric state is consistently populated under physiological conditions suggests that this shift would be fast and effective.

Nonetheless, it is interesting to follow the interaction between the p--LES and the HIV--1--PR, starting from the protein in its dimeric native conformation. For this purpose, we have performed $10^{10}$ MCS simulations of a system composed of the dimer and three p--S$_2$ peptides at $T=2.5$ kJ/mol. Due to its large size, the equilibration of the system is computationally quite demanding (3 weeks on a Xeon workstation).  The population $p_N$ of the native dimeric state (using the same definition as above) is $0.05$, to be compared with $0.304$ for the case in which the isolated dimer is evolved starting again from the native conformation. The detailed values for the populations of the various states are given in Table \ref{tab_ultima}. The large errors ascribed to these numbers reflect the computational difficulties in equilibrating the system.

\begin{figure}
\centerline{\psfig{file=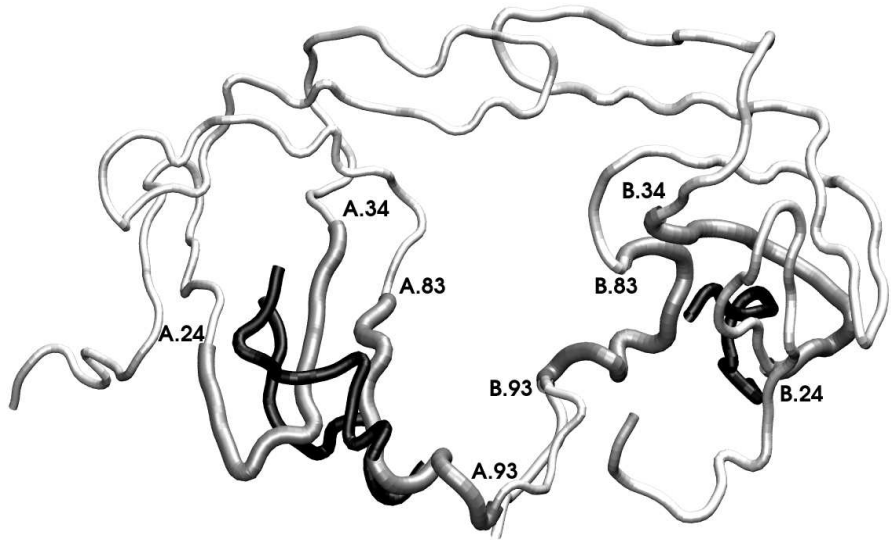,width=12cm}}
\caption{A snapshot of the simulation of the dimer with 3 p--LES, starting from the native conformation. The LES of the protein are indicated with a thicker grey tube, while the p--LES are highlighted in black.}
\label{fig_dimer}
\end{figure}

A snapshot of the result of the three p--S$_8$ plus dimer simulation is shown in Fig. \ref{fig_dimer}. It is seen that the p--S$_8$ peptides are able to bind to the protein, blocking the way of the native S$_8$ LES to dock the S$_2$ LES, thus disrupting the folding nucleus. This situation is similar to that shown in Fig. \ref{snap} (monomer plus three p--S$_8$ peptides). In the present case, when the p--S$_8$ enters the protein, the RMSD of the associated monomer is increased to a value $\approx 14\AA$, which implies that the native conformation is lost. 

\section{Effects of mutations in HIV--1 PR sequence}

In order to assess the effect  mutations of the protease have on the effectiveness of the p--S$_8$ as inhibitor, we have performed a number of simulations of mutated protease with and without p--S$_8$. Within the framework of the present model a mutation on a given site is made operative by switching off all the native interactions made by that site in the wild--type sequence, treating them as if they were non--native (cf. Eq. (\ref{potential})). We have applied this procedure to a number of sites which are known to be mutated by the virus to escape drugs (e.g., 19, 37, 63, 67, 72 and 95). Also to sites which belong to LES (31, 33 and 85) or which interact with a LES (68).

\begin{figure}
\centerline{\psfig{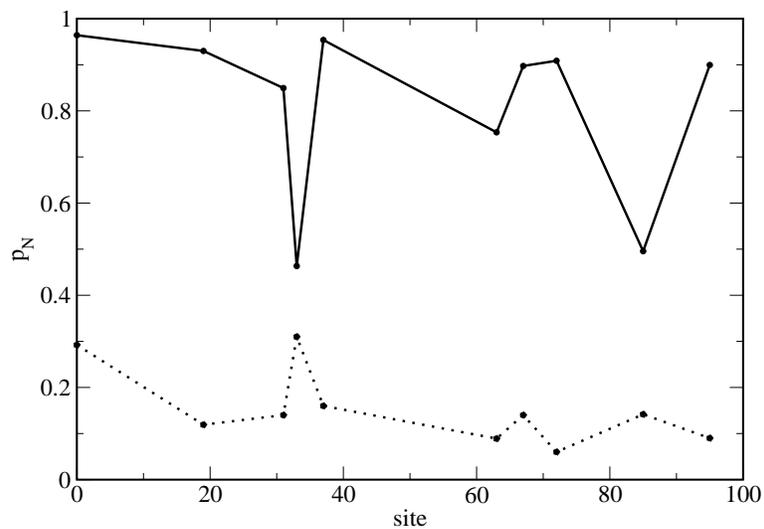}}
\caption{The effect of mutations on a number of sites of the monomer (x--axis) on the stability $p_N$ of the native state (y--axis). The solid curve indicate the stability of the monomer alone ($T=2.5$kJ/mol), while the dotted curve refers to the case of the monomer plus 3 p--S$_8$ peptides. The point drawn at abscissa zero indicates the wild type sequence.}
\label{fig_mut}
\end{figure}

In Fig. \ref{fig_mut} we display the population of the native state $p_N$ for the mutated monomeric protease (continuous curve) and for the system composed of the mutated protease and three p--S$_8$ (dotted curve). All mutations except those on sites 85 and 33 have little effect on the stability of the protein.  On the other hand, the denaturing effect of the p--S$_8$ is fully retained, as expected from the fact that the interaction between p--S$_8$ and the monomer is unchanged. In fact, the interaction between p--S$_8$ and the monomer are the same as those which stabilize the monomer itself. Consequently, the mantained stability of the mutated protein is a proof of the mantained affinity between p--S$_8$ and the mutated protein.

The mutation on site 33 causes a consistent destabilization of the protease, due to the fact that it is a hot site belonging to a LES (S$_2$), site where a large fraction of the stabilization energy of the protein is concentrated. In this case the affinity of the p--S$_8$ to the protease is also diminished and consequently its destabilizing effect is greatly reduced. In any case the net effect of this mutation is that the protein becomes quite unstable.

\section{Conclusions}

We have identified fragments 24--34, 83--93 and 75--78 as the local elementary structures which guide the folding and are responsible for the stability of HIV--1--PR. Peptides with the same sequence of the fragment 83--93 have shown good inhibitory properties, causing a consistent unfolding of its native state. It is not likely that HIV--1--PR can develop resistance against these peptides. To do so, the protease should mutate amino acids which play an important role in its folding. The strategy presented in this paper seems to solve the two major problems encountered in drug desing: optimization and resistance. Due to the universality of the approach, it is  suggested that this kind of folding--inhibitor drugs can be designed and used in connection with other target proteins.


\appendix
\section{Nature of the HIV--Pr dimer} \label{appA}

At low pH, calorimetry experiments have shown \cite{todd98} that there is a single transition at $T=59$ C (pH 3.4, 25 $\mu$M protein, 100 mM NaCl) between the dimeric native state and a monomeric unfolded state. This means that, for example, at T=25 C there is essentially not monomeric protease in solution. At this temperature, the stabilization energy of the dimer is about 10 kcal/mol, but each pair of isolated monomers is unstable (the stabilization free energy being $\approx -13$ kcal/mol), the stabilization energy coming from the interface ($\approx +20$ kcal/mol).

The analysis of the distribution of stabilization energy among the residues at the interface indicate that this is concentrated in few "hot" spots, namely 1, 3, 5 and 95--99. This behaviour is somewhat odd for a two--state dimer, where typically the stabilization energy is spread out uniformly on the interface \cite{dimer}. Note, however, that most likely the evolution of the HIV protease has mostly taken place in the cytoplasm, which is neutral, and consequently one should compare its evolutionary properties with its stability features at higher pH.

Increasing the value of pH, acid residues acquire a negative charge. In particular, the pair of D25 which lie close on the interface repell each other through Coulomb force. The overall effect is to increase the dissociation constant (measured by sedimentation equilibrium experiments) which assumes the value $k_D=5.8 \mu$M at pH 7 (and $T=4$C \cite{xie}, further increasing at higher temperatures). The dissociation of dimer is accompanied with a decrease in the internal structure of the monomers, as testified by the fact that CD experiments highlight a decrease of about 26\% of the beta--structure of the protein \cite{xie}. Consequently, one expects a detectable ratio of folded monomers in solution. For instance, according to a picture where a loss of 26\% structure means that 26\% of monomers have no structure at all, the ratio of folded monomers is 18 \% (of course, this is an idealized situation. In a more realistic scenario, a larger ratio of monomers are partially destabilized, e.g. 52\% of monomers loose half of the structure). 

One should anyway note that sedimentation equilibrium results are apparently contraddicted by fluorescence essays. The intensity of fluorescence at equilibrium at different values of pH displays maximum values in the region between pH 4.5 and pH 8, decreasing both at acidic and basic values of pH \cite{szeltner96,grant92}. Moreover, equilibrium experiments where the fluorescence is recorded upon addition of urea at constant pH and the concentration of urea corresponding to the midpoint in the change in fluorescence \cite{todd98} show that the midpoint concentration of urea increases if pH is increased from 4 to 5.5. These results seem to contraddict the results discussed above, suggesting that the protease is more stable at neutral pH. They also seem to contraddict kinetic fluorescence measurements, where the kinetics of fluorescence emission upon addition of urea is recorded at different pH. The "unfolding rates" thus obtained behave in a specular way as compared to equilibrium measurement, displaying low rates (i.e., "more stable") between pH 4 and 7, and increasing (i.e., "less stable") at the extremes \cite{szeltner96}.

A summary of the dissociation constants obtained with different tecniques and under different conditions is presented in Table \ref{table_kd}. In considering these results, one should remember that each monomer of the protease has two TRP residues, at position 6 at the interface and at position 42 on the surface, at the rear part of the flap. Consequently, not only fluorescence intensity is not able to distinguish between unfolding of the monomer and dissociation of the dimer, but more subtle processes such as opening and closing of the flaps can affect the recorded signal. As a consequence, we consider more reliable the results obtained in sedimentation equilibrium experiments, which is the only direct inspection of the monomer/dimer character of the solution.

\section{Dynamic simulations} \label{appB}

Metropolis Monte Carlo simulations are meant to sample the conformational space of a system in order to investigate thermodynamical quantities at equilibrium. Nonetheless, it has been shown that, if the elementary movement of the chain is small enough, the algorithm describes in an approximated way the dynamics of the system \cite{rey}, solving effectvely the Langevin equations associated to the system \cite{nippon}. In order to transform MC steps into seconds, we have calculated the diffusion coefficient of the centre of mass of the protein monomer, and compared it with the one obtained by Stokes' approximation, describing diffusion of a spherical object of radius $30\AA$ in water at room temperature. The resulting relation is 1 MC step equal to $10^{-13}$s.



\clearpage
\newpage

\begin{table}
\begin{tabular}{|c|c|c|c|}
\hline
contact & $B_{ij}$ & $p_\infty$ & $\tau$ \\\hline
25-28 & -5.4 & 0.77 & $2\cdot 10^{-10}$ \\
87-90 & -9.8 & 0.99 & $4\cdot 10^{-10}$ \\
31-89 & -6.1 & 0.92 & $2.6\cdot 10^{-7}$ \\
23-85 & -4.9 & 0.86 & $3.6\cdot 10^{-7}$\\
62-74 & -2.4 & 0.51 & $1.0\cdot 10^{-6}$ \\
12-66 & -1.5 & 0.51 & $1.2\cdot 10^{-6}$ \\
\hline
\end{tabular}
\caption{The dynamics of some native contacts of the protein. $B_{ij}$ is the interaction energy expressed in kJ/mol, $p_\infty$ is the asymptotic stability and $\tau$ is the average formation time of the contact in seconds. }
\label{table_dyn}
\end{table}

\clearpage
\begin{table}
\tiny
\begin{tabular}{|l|l|l|l|l|l|l|l|l|l|l|l|} \hline
wt	& mut	& PAM	& wt	& mut	& PAM	&wt	& mut	& PAM	&wt	& mut	& PAM	\\\hline
P1	&	&	&T26	&	&	&G51	& 	&	&L76	& V	&2	\\  
Q2	&	&	&G27	&	&	&G52	&	&	&V77	& I	&4	\\
I3	& V	&4	&A28	&	&	&F53	& L	&2	&G78	& 	&	\\
T4	&	&	&D29	&	&	&{\bf I54}& VMLT&0	&P79	& A	&1	\\
L5	&	&	&D30	& N	&2	&{\bf K55}& RH	&0	&T80	&	&	\\
W6	&	&	&T31 	&	&	&V56	&	&	&{\bf P81}& T	&0	\\
Q7	&	&	&V32	& I	&4	&R57 	& K	&3	&{\bf V82}& TAFIS&-1	\\
R8	& KQL	&1	&L33	& FVI	&2	&Q58	& E	&2	&N83	&	&	\\
P9	&	&	&{\bf E34}& DQANG&0	&Y59	&	&	&I84	& V	&4	\\
{\bf L10}& FIRV	&-3	&{\bf E35}& DG	&0	&D60	& E	&3	&I85	& V	&4	\\
V11	& IL	&2	&M36	& ILV	&2	&Q61	& ENH	&1	&G86	&	&	\\
{\bf T12}&SPAEIKN&0	&{\bf S37}&DSTEKHC&-4	&I62	& V	&4	&R87	& K	&	\\
I13	& V	&4	&L38	& F	&2	&{\bf L63}&PSTACQH&-6	&N88	& DS	&1	\\
K14	& R	&3	&{\bf P39}& SQT	&0	&I64	& VLM	&2	&L89	& MVI	&2	\\
I15	& V	&4	&G40	&	&	&E65	& D	&3	&L90	& M	&4	\\
{\bf G16}& EA	&0	&{\bf R41}& KN	&0	&I66	& FV	&1	&T91	& 	&	\\
{\bf G17}& E	&0	&W42	&	&	&{\bf C67}& FS	&-4	&Q92	& KR	&1	\\
Q18	& H	&3	&{\bf K43}& RT	&0	&G68	&	&	&I93	& L	&2	\\
{\bf L19}& ITVQP&-3	&P44	&	&	&{\bf H69}& KQYRN&0	&G94	&	&	\\
{\bf K20}& MRTIV&-2	&{\bf K45}& IRN	&-2	&{\bf K70}& RTE	&0	&{bf A95}& SF	&-3	\\
E21	&	&	&M46	& FILV	&0	&{\bf A71}& TVI	&-1	&T96 	&	&	\\
A22	&	&	&I47	& V	&4	&{\bf I72}&VTLMER&-2	&L97	& V	&2	\\
L23	& I	&2	&{bf G48}& V	&-1	&G73	& STCA	&1	&N98	&	&	\\
L24	& IVF	&2	&G49	&	&	&{\bf T74}& SAP	&0	&F99	&	&	\\
D25	&	&	&I50	& VL	&2	&V75	& I	&4	& 	&	&	\\
\hline
\end{tabular}
\caption{The observed mutations in proteases reported in ref. \protect\cite{shafer}. For each residue of the wild--type sequence (wt) are listed the mutations observed in treated or untreated patients (mut) and the lowest PAM250 score (see text) among the mutations. In bold the sites which undergo non--conservative (i.e., non--positive) mutations.}
\label{table_mut}
\end{table}

\begin{table}
\begin{tabular}{|c|c|c|c|}
\hline
 & S$_1$+S$_6$ & S$_2$+S$_7$+S$_8$ & S$_3$+S$_4$+S$_5$\\\hline
 S$_1$+S$_6$ & -67.2 & -72.6 & -19.2 \\\hline
 S$_2$+S$_7$+S$_8$ & -72.6 & -223.6 & -117.2 \\\hline
 S$_3$+S$_4$+S$_5$ & -19.2 & -117.2 & -144.1 \\\hline
\end{tabular}
\caption{The all atom molecular dynamics energy map of the eight amino acid chain segments expressed in terms of the folding nucleus (FN: S$_2$+S$_7$+S$_8$), the flap region (S$_3$+S$_4$+S$_5$) and of S$_1$+S$_6$. }
\label{table1}
\end{table}

\begin{table}
\begin{tabular}{|c|l|l|l|l|}\hline
$n_p$ & $N_2$(\%) & $2N$(\%) & $U$(\%) & $U_2$(\%) \\\hline
0 &  $30.4\pm 29.1$ & $68.4\pm 29.0$ & $1.1\pm 0.1$ & $0.1\pm 0.1$ \\\hline
3 &  $0.05\pm 0.05$ & $31.3\pm 16.2$ & $68.6\pm 21.1$ & $0.05\pm 0.05$ \\\hline
\end{tabular}
\caption{The equilibrium population $p_N$ ($q_E>0.7$, $RMSD<10\AA$) of the different states of the HIV--1--PR dimer in the presence of $n_p$ p--S$_8$ peptides at $T=2.5$kJ/mol: $N_2$) native dimeric state, $2N$) monomers in the native state but not dimerized, $U$) monomers unfolded and separated, $U_2$) monomers unfolded but native interactions at the dimer interface formed.}
\label{tab_ultima}
\end{table}

\begin{table}
\begin{tabular}{|c|c|c|c|c|c|} \hline
pH & T [C] & NaCl & $k_D$ & method & reference \\\hline
3.4 & 25 & 100 mM & 17 nM & fluorescence & \protect\cite{todd98} \\
4.5 & 4 & 0 & $<$100 nM & sedim. eq. & \protect\cite{xie} \\
4.5 & 9 & 0 & 28 $\mu$M & sedim. eq. (HIV--2)& \protect\cite{holzman91} \\
5 & 4 & 0.2 M& $<$10 nM & sedim. eq. & \protect\cite{grant92} \\
5 & 25 & 100 mM & 23 pM & fluorescence & \protect\cite{todd98} \\
5 & 37 & 1 M & 3.6 nM & activity & \protect\cite{zhang91} \\ 
7 & 4 & 0 & 5.8 $\mu$M & sedim. eq. & \protect\cite{xie} \\
7 & 25 & 1 M & 50 nM & activity & \protect\cite{cheng90} \\
7.5 & 9 & 0 & 87 $\mu$M & sedim. eq. (HIV--2) & \protect\cite{holzman91} \\
\hline
\end{tabular}
\caption{The values of dissociation constant $k_D$ if the dimer, calculated at different conditions.}
\label{table_kd}
\end{table}


\end{document}